\documentclass[aps,prl,superscriptaddress,twocolumn]{revtex4-1}
\usepackage{graphicx}  % needed for figures
\usepackage{dcolumn}   % needed for some tables
\usepackage{amsmath}
\usepackage{braket} %Provides macros to typeset bra-ket notation
\usepackage{lineno}
\usepackage{xcolor}
\usepackage{amssymb}
\usepackage{ulem}
\newcommand{\RNum}[1]{\uppercase\expandafter{\romannumeral #1\relax}}

\begin{document}

\title{Honeycomb-Lattice Mott insulator on Tantalum Disulphide}

\author{Jinwon Lee}
	\affiliation{Center for Artificial Low Dimensional Electronic Systems, Institute for Basic Science (IBS), Pohang 37673, Republic of Korea}
	\affiliation{Department of Physics, Pohang University of Science and Technology, Pohang 37673, Republic of Korea}
	
\author{Kyung-Hwan Jin}
    \affiliation{Center for Artificial Low Dimensional Electronic Systems, Institute for Basic Science (IBS), Pohang 37673, Republic of Korea}
    \affiliation{Department of Materials Science and Engineering, University of Utah, Salt Lake City, Utah 84112, United States}
    
\author{Andrei Catuneanu}
    \affiliation{Department of Physics, University of Toronto, Ontario M5S 1A7, Canada}
    
\author{Ara Go}
    \affiliation{Center for Theoretical Physics of Complex Systems, Institute for Basic Science (IBS), Daejeon 34126, Republic of Korea}
    
\author{Jiwon Jung}
    \affiliation{Center for Artificial Low Dimensional Electronic Systems, Institute for Basic Science (IBS), Pohang 37673, Republic of Korea}
	\affiliation{Department of Physics, Pohang University of Science and Technology, Pohang 37673, Republic of Korea}

\author{Choongjae Won}
    \affiliation{Laboratory for Pohang Emergent Materials, Pohang University of Science and Technology, Pohang 37673, Republic of Korea}
    \affiliation{Rutgers Center for Emergent Materials and Department of Physics and Astronomy, Rutgers University, Piscataway, NJ 08854, United States}

\author{Sang-Wook Cheong}
    \affiliation{Laboratory for Pohang Emergent Materials, Pohang University of Science and Technology, Pohang 37673, Republic of Korea}
    \affiliation{Rutgers Center for Emergent Materials and Department of Physics and Astronomy, Rutgers University, Piscataway, NJ 08854, United States}

\author{Jaeyoung Kim}
    \affiliation{Center for Artificial Low Dimensional Electronic Systems, Institute for Basic Science (IBS), Pohang 37673, Republic of Korea}

\author{Feng Liu}
    \affiliation{Department of Materials Science and Engineering, University of Utah, Salt Lake City, Utah 84112, United States}

\author{Hae-Young Kee}
    \affiliation{Department of Physics, University of Toronto, Ontario M5S 1A7, Canada}
    \affiliation{Canadian Institute for Advanced Research, CIFAR Program in Quantum Materials, Toronto, ON M5G 1M1, Canada}

\author{Han Woong Yeom}
	\email{yeom@postech.ac.kr}
	\affiliation{Center for Artificial Low Dimensional Electronic Systems, Institute for Basic Science (IBS), Pohang 37673, Republic of Korea}
	\affiliation{Department of Physics, Pohang University of Science and Technology, Pohang 37673, Republic of Korea}

\begin{abstract}
    Effects of electron many-body interactions amplify in an electronic system with a narrow bandwidth opening a way to exotic physics.
    A narrow band in a two-dimensional (2D) honeycomb lattice is particularly intriguing as combined with Dirac bands and topological properties but the material realization of a strongly interacting honeycomb lattice described by the Kane-Mele-Hubbard model has not been identified.
    Here we report a novel approach to realize a 2D honeycomb-lattice narrow-band system with strongly interacting 5\textit{d} electrons.
    We engineer a well-known triangular lattice 2D Mott insulator 1\textit{T}-TaS$_2$ into a honeycomb lattice utilizing an adsorbate superstructure.
    Potassium (K) adatoms at an optimum coverage deplete one-third of the unpaired \textit{d} electrons and the remaining electrons form a honeycomb lattice with a very small hopping.
    \textit{Ab initio} calculations show extremely narrow Z$_2$ topological bands mimicking the Kane-Mele model.
    Electron spectroscopy detects an order of magnitude bigger charge gap confirming the substantial electron correlation as confirmed by dynamical mean field theory.
    It could be the first artificial Mott insulator with a finite spin Chern number.
 \end{abstract}

\maketitle

%-----Introduction-------------------------------------------

\textit{Introduction}---An electronic system with an extremely narrow bandwidth has recently attracted huge renewed interests, as it offers a playground to discover exotic quantum phases due to much stronger electron interaction relative to its kinetic energy~\cite{Bergholtz2013,Derzhko2015}.  
   A narrow band in a two-dimensional (2D) honeycomb lattice is particularly intriguing,
   as it hosts Dirac electrons with a linear low-energy dispersion as in graphene~\cite{Novoselov2005,Zhang2005}, and when the spin-orbit coupling is introduced, 
   a gap opens at Dirac points, leading to a quantum spin Hall state~\cite{Kane-Mele2005}.
   A sufficiently strong on-site Coulomb repulsion (\textit{U}) can induce a Mott insulator phase, as described by the Kane-Mele-Hubbard model~\cite{Hohenadler2012}, and various exotic phases have been proposed to emerge such as topologically non-trivial states~\cite{Raghu2008,Laubach2014}, quantum spin liquids~\cite{Lee2005}, charge density waves~\cite{Raghu2008}, and superconductivity~\cite{Uchoa2007} at or near the half filling, and magnetic Chern insulator at a quarter filling~\cite{Mishra2018}. 
   However, the realization of a 2D honeycomb lattice with strong electron correlation has been challenging since Dirac electrons usually have high kinetic energy.
    A breakthrough in this challenge was recently achieved by making the Dirac band very flat by twisting two graphene sheets by a magic angle~\cite{Cao2018,Cao2018-2}.
   In this system, a correlated insulator and superconductivity~\cite{Cao2018,Cao2018-2} and more recently a ferromagnetic insulator with a finite Hall conductivity were reported~\cite{Kang2019,Seo2019,Sharpe2019,Serlin2020,Repellin2020}.
   Despite the rapid progress of the research on twisted-bilayer-graphene with very small Coulomb energy, the material realization of a Kane-Mele-Hubbard system has been elusive.
   
%&&&&&&&&&&&&&&&&&&&&&&&&&&&&&&&&&&&&&&&&&&&&&&&&&&&&&&&&&&&&
\begin{figure*}[t]
\includegraphics[width=15.8 cm]{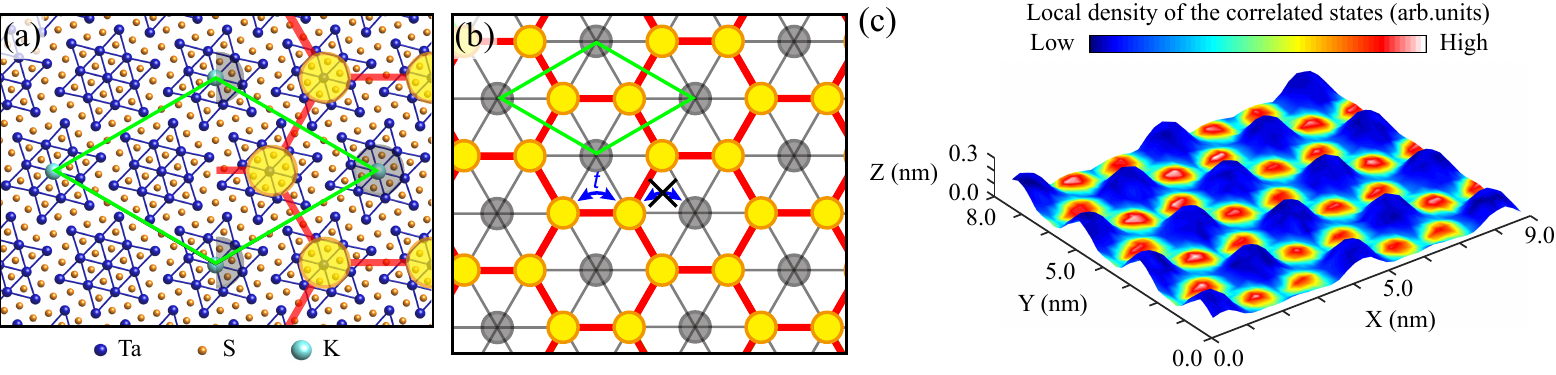}
\caption{\label{Fig01}
Honeycomb-lattice correlated phase arising from K adsorption on 1{\textit {T}}-TaS$_2$.
(a)~Atomic structure of K-adsorbed 1{\textit {T}}-TaS$_2$. Blue stars indicate CDW superstructures (David stars) and a green rhombus is a unit cell of K-adsorbed 1{\textit {T}}-TaS$_2$.
(b)~A schematic for the honeycomb-lattice correlated phase. Every circle indicates each David star. K-adsorption sites (fully filled states) and bare David stars (half-filled states) are illustrated with gray and yellow color, respectively.
Note that a half of (a) is overlaid with the schematic of (b).
(c)~3D illustration of the STM topography of K-adsorbed 1{\textit {T}}-TaS$_2$ (height map) with the local density of the upper Hubbard states (UHS, color map).}
\end{figure*}
%&&&&&&&&&&&&&&&&&&&&&&&&&&&&&&&&&&&&&&&&&&&&&&&&&&&&&&&&&&&&

    In this letter, we report a unique approach to realize a 2D honeycomb-lattice narrow-band system with a strongly-interacting 5$d$ transition metal.
    We start from a triangular lattice with a substantial Coulomb interaction and device a way to manipulate the electron hopping into the honeycomb symmetry. 
    The system is based on a well-known 2D Mott insulator 1$T$-TaS$_2$ and akali metal adsorbates are utilized as the lattice modifier.
    1{\textit {T}}-polytype TaS$_2$ undergoes a series of spontaneous transitions into various charge-density-wave (CDW) phases~\cite{Wilson1975}.
    The common unit cell of the CDW phases, known as a David-star cluster, is composed of 13 distorted Ta atoms in a $\sqrt{13}\times\sqrt{13}$ triangular superlattice (Fig.~1(a)). 
    Each cluster contains 13 Ta 5$d$ electrons, twelve of which pair to form insulating CDW bands and to leave one unpaired electron in a half-filled metallic band. 
    The latter is very flat since the hopping to neighboring David stars is limited by a large distance of $\sim$~1.2\,nm between them.
    As a result, a moderate electron correlation $U$ introduces a Mott gap of about 0.2-0.4~eV at low temperature~\cite{Barker1975,Gasparov2002,Zwick1998,Clerc2006,Perfetti2008,Sato2014,Ritschel2015,Lutsyk2018,Kim1994,Kim1996,Cho2015,Cho2016,Ma2016,Cho2017,Butler2020}.
    The superconductivity emerging from this insulting state by pressure~\cite{Sipos2008} or chemical doping~\cite{Xu2010,Li2012,Liu2013} was reported and a quantum spin liquid phase was suggested recently ~\cite{Law2017,Ribak2017,Klanjsek2017}.

    As shown in Fig.~1(b), one can convert a triangular lattice into a honeycomb by eliminating 1/3 of the lattice sites periodically. 
    Fig.~1(c) shows a scanning tunneling microscopy (STM) topograph combined with a map of local density of states (LDOS, represented by color) for a 1{\textit {T}}-TaS$_2$ surface with potassium (K) adatoms of an optimal coverage.
    At this coverage, K atoms (highly protruded in topograph and blue in the LDOS map) sit selectively at the center of a David-star cluster and form a $(\sqrt{3}\times\sqrt{3})R30^{\circ}$ superstructure with respect to the CDW lattice (green rhombuses in Fig.~1(b)).
    K atoms indeed deplete the unpaired $d$ electron states within a David-star cluster on which it sits as shown in the LDOS map (Fig.~1(c)).
    The remaining unpaired electrons in neighboring clusters are little affected by K adatoms and form a honeycomb lattice as illustrated in Figs.~1(b) and 1(c). 
    The Dirac electrons with extremely narrow dispersions and the correlation gap in this system will robustly be justified below.

\textit{Methods}---1\textit{T}-TaS$_2$ single crystal was cleaved in high vaccum to obtain clean surfaces and the K deposition was conducted in ultra high vacuum ($1\times 10^{-10}\,torr$) and at room temperature. STM measuremets were performed at 4.4~K in the constant-current mode with a sample bias of -600~meV. A lock-in technique was used in scanning tunneling spectroscopy with a modulation of 1~kHz.
    The single-particle band structures and their topological properties were investigated within the framework of the density functional theory (DFT) and tight binding calculations~\cite{supplement} and the electron correlation was considered in the dynamical mean field theory (DMFT) calculation~\cite{supplement}.

\textit{Honeycomb lattice in K-adsorbed 1{\textit {T}}-TaS$_2$}---The STM image in Fig.~2(a) resolves David-star clusters as medium-contrast protrusions, which is consistent with the previous studies~\cite{Giambattista1990,Cho2015}.
    The exceptional features, the prominently bright protrusions, are David-star clusters with one K adatom in each cluster.
    At the optimized coverage, one third of the CDW unit cells are rather regularly occupied by K atoms.
    The $\sqrt{3}\times\sqrt{3}$ superstructure of adatoms with respect to the CDW unit cell (the unit cell of the green rhombus in Fig.~2(a)) can be confirmed by the Fourier-transformed STM image of a wide area in Fig.~2(g) (see SFig.~1 in the Supplemental Material~\cite{supplement}).
    %The adatom ordering comes from the dipolar repulsion between positively ionized K atoms by donating electrons into the substrate. 
    This will be made clear below.

    One can notice that David-star clusters with or without K adatoms have totally different electronic states in the scanning tunneling spectra (the normalized $dI/dV$ curves) as shown in Fig.~2(h).
    Though both spectra show the charge gap, one can clearly see the prominent peaks at -\,90 and 150\,mV on a bare David-star cluster (red data), which are absent in the clusters with K adatoms (green data). 
    Note that the gaps on two different sites have totally different origin.
    A bare David star has odd number of electrons and its energy gap must originate from a correlated insulating phase as detailed below.
    The spectral features on a bare David-star cluster are consistent with those of the pristine 1\textit{T}-TaS$_2$, which is established as an Mott insulator (see SFig.~2 in the Supplemental Material~\cite{supplement}).
    On the other hand, a David star with a K adsorbate has even number of electrons due to the electron donation from K, leading to a simple one-electron gap.
    The local suppression of the presumed upper and lower Hubbard states (UHS and LHS) by K adatoms is well visualized in the corresponding LDOS maps, or normalized $dI/dV$ maps, of Figs.~2(d) and 2(e), respectively.
    These LDOS maps further tell us that the correlated states seem to survive in a honeycomb hopping lattice.
    Note also that the LDOS maps corresponding to CDW bands (Figs.~2(b) and 2(f)) are the same as those in the adsorbate-free area or a pristine sample (see SFig.~2 in the Supplemental Material~\cite{supplement}), telling us that the backbone CDW structure is intact even after the K adsorption.

%&&&&&&&&&&&&&&&&&&&&&&&&&&&&&&&&&&&&&&&&&&&&&&&&&&&&&&&&&&&&
\begin{figure*}[t]
\includegraphics[width=17.6 cm]{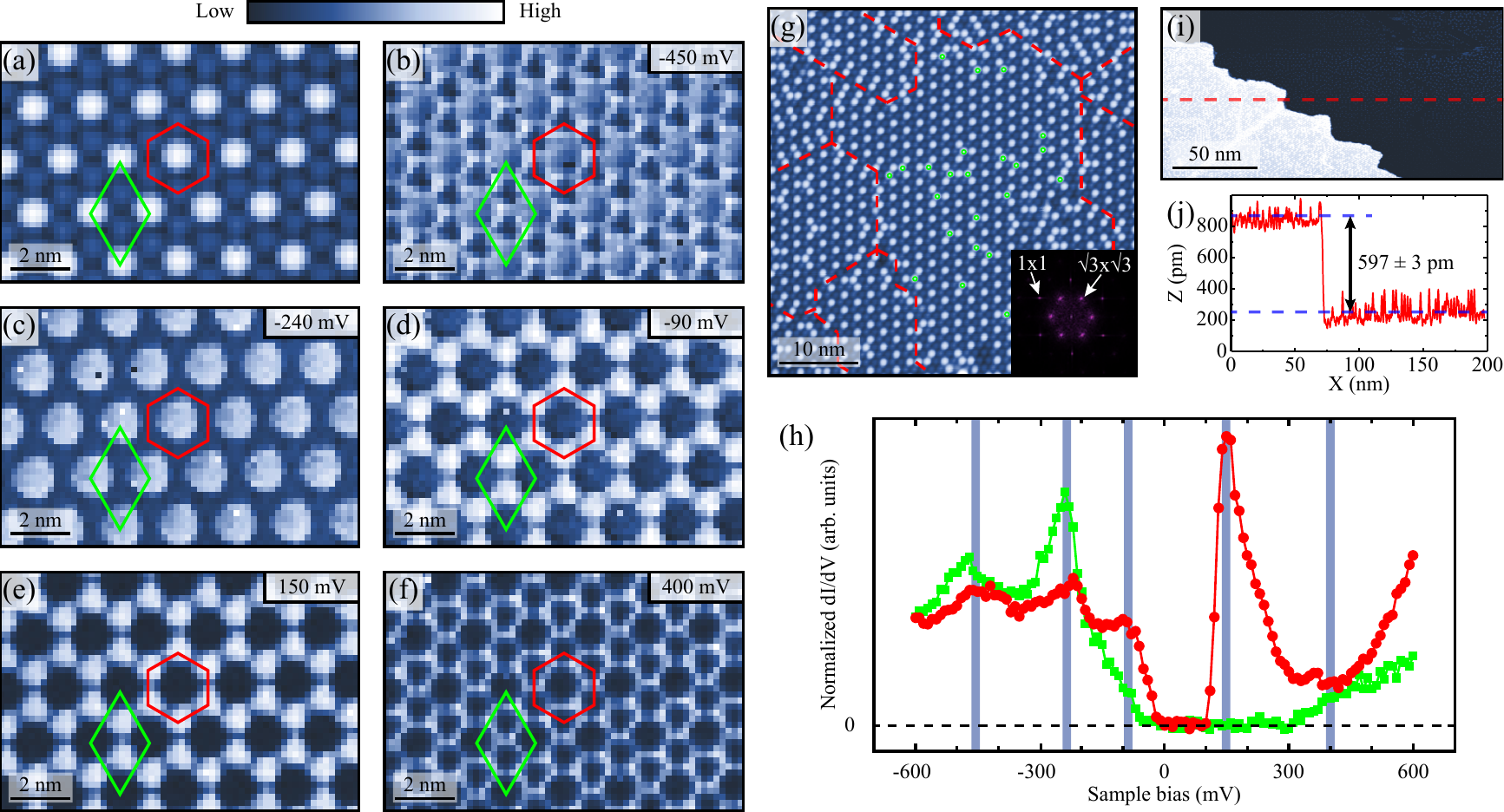}
\caption{\label{Fig02}
Local electronic states of K-adsorbed 1{\textit {T}}-TaS$_2$.
(a)~STM topographic image of the K-adsorbed 1{\textit {T}}-TaS$_2$ surface.
(b)-(f)~Corresponding normalized $dI/dV$ images for $V$\,=\,-\,450, -\,240, -\,90, and 150, and 400\,mV (see vertical bars in (h)). Note that the features at -\,90 and 150\,mV correspond to correlated Dirac states.
(g)~A wide-area STM image of 1{\textit {T}}-TaS$_2$ and its Fourier transform in the inset.
$1\times 1$ and  $\sqrt{3}\times\sqrt{3}$ stand for the structures with respect to the CDW unit cell.
(h)~Normalized $(dI/dV)$ spectra at fixed positions. Green and red plots are obtained on the K-adsorption site (green rhombus) and the bare David star (red hexagon), respectively.
(i)~STM image at a step edge in 1{\textit {T}}-TaS$_2$ after K deposition.
(j) Height ($Z$) profile along the red dashed line in (i).
}
\end{figure*}
%&&&&&&&&&&&&&&&&&&&&&&&&&&&&&&&&&&&&&&&&&&&&&&&&&&&&&&&&&&&&
    
    By passing, we comment further on the adsorption behavior of K atoms. 
    First of all, K atoms are not intercalated but adsorbed on top of the surface S layer as confirmed by the layer spacing (597~pm) (Figs.~2(i) and 2(j))~\cite{Schmidt2006}.
    DFT calculates the adsorption energy of a K atom on various possible sites within a CDW unit cell (see Supplementary Note~3, STable~\RNum{1}, and SFig.~3 in the Supplemental Material~\cite{supplement}) and finds
    the center of the David star to be the most energetically favorable.
    The 4$s$ valence electron of a K adatom is almost completely transferred to the substrate to make the adatom ionic (Fig.~3(a)). It is highly localized on a single David-star cluster, leaving neighboring clusters intact and forming a local dipole.
    The dipole-dipole interaction explains the repulsion between adatoms to form the ordered superstructure with a domain size upto 30$\times$50~nm$^2$ (Fig.~2(g)). 
    Given the short-range repulsion, K atoms adsorb rather randomly avoiding the nearest-neighbor David stars below the optimum coverage (see SFig.~4 in the Supplemental Material~\cite{supplement} for more details of the growth behavior).
    These calculations agree well with the STM and STS observations.

%&&&&&&&&&&&&&&&&&&&&&&&&&&&&&&&&&&&&&&&&&&&&&&&&&&&&&&&&&&&&
\begin{figure*}[t]
\includegraphics[width=17.2 cm]{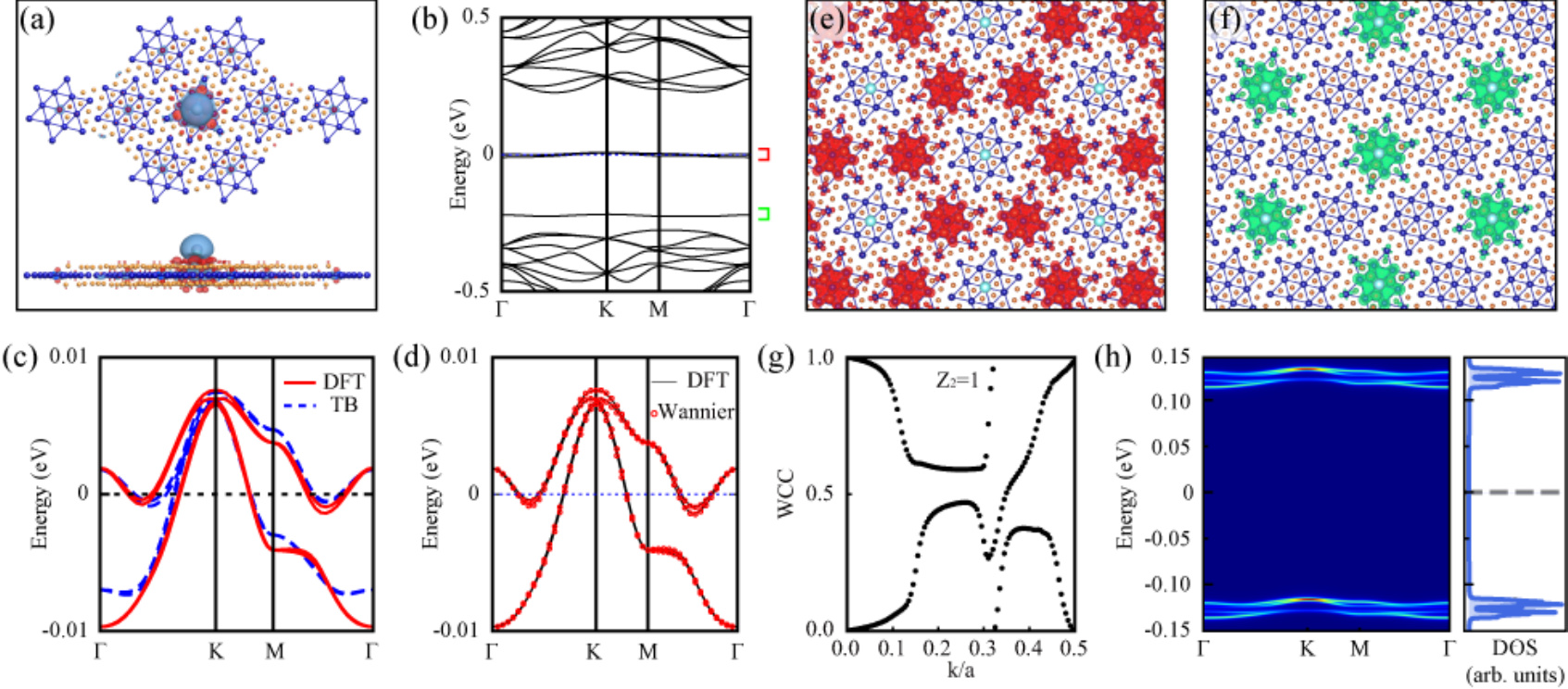}
\caption{\label{Fig03}
Electronic band structure of K-adsorbed 1{\textit {T}}-TaS$_2$.
(a)~Charge transfer from the K adsorbate. Red and blue colors represent accumulation and depletion of electron, respectively. Note that electron transferred from K adatom is highly localized at the adsorption site.
(b)~DFT band structure with a paramagnetic constraint.
(c) and (d)~Enlargement of the band structure near the Fermi level with the tight binding (TB) fitting, and the maximally localized Wannier functions fitting, respectively.
(e)~A spatial distribution of the metallic Dirac bands (marked with red in (b)), and (g)~that of a fully occupied band at $\sim$~-0.2~eV (marked with green in (b)).
(g)~Evolution of Wannier charge centers by varying the pumping parameter $k$. The evolution curves have an odd-number crossing with respect to an arbitrary horizontal line.
(h)~Band structure after considering the electron correlation and the electronic density of states (DOS). The metallic Dirac bands at the Fermi level split with a large charge gap. The extra gap like feature below each Dirac point is an artifact of the present single-site DMFT calculation.
}
\end{figure*}
%&&&&&&&&&&&&&&&&&&&&&&&&&&&&&&&&&&&&&&&&&&&&&&&&&&&&&&&&&&&&

\textit{Electronic structures and topology}---The DFT calculation for 1{\textit {T}}-TaS$_2$ with K adatoms in a $(\sqrt{3}\times\sqrt{3})R30^{\circ}$ superstructure predicts two extremely narrow bands at the Fermi energy within the CDW gap as shown in Fig.~3(b).
    While the two bands are expected from effectively two orphan electrons in a honeycomb supercell with two David-star clusters, the narrow bandwidth with almost linear crossing shown in Figs.~3(c) and 3(d) is notable.
    The small bandwidth of $\sim$~20~meV is due to the $d_{z^2}$ orbital character of those orphan electrons and the huge size of the supercell (see SFig.~5 in the Supplemental Material~\cite{supplement}).
    The direct hopping integrals between $d$ orbitals in the honeycomb plane are significantly reduced, and the hopping paths are mainly achieved via $p$ orbitals of S atoms.
    Given that the inversion symmetry is broken, a finite spin-orbit coupling (SOC) including the Rashba effect splits these bands into four with a tiny gap of $\sim$~0.5~meV at K point.
    Due to time-reversal symmetry, the bands are degenerate at the time-reversal-invariant-momentum points of $\Gamma$ and M.
    In addition, a fully occupied band below the Fermi level appears at adsorption sites (marked with a green color in Fig.~3(b)) as originates from the Ta~5$d$ level doped with K 4$s$ electrons.
    Its extremely narrow dispersion is due to the strong localization of donated electrons (Fig.~3(a)).
    The spatial distributions of two distinct types of electrons are visualized in Figs.~3(e) and 3(f), nicely matching the experimental results (Figs.~2(c)-2(e)).
   
    We investigate the topology of these flat bands by a tight binding model.
    Indeed, the tight binding model obtained through the Wannier function~\cite{Mostofi2014}, by projecting the Bloch states from the DFT calculation onto the Ta $d_{z^2}$ orbital (see Note~4, SFigs.~5,~and~6 in the Supplemental Material~\cite{supplement}), becomes a characteristic Kane-Mele type Hamiltonian.
    \begin{equation}
    \begin{aligned}
    H = & \sum_{i, j,s}  t_i c^{\dagger}_{i,s} c_{j,s} +i \lambda_R \sum_{\langle ij \rangle} c^\dagger_{i,s} ({\vec \sigma} \times {\hat d}_{ij})^z_{s,s'} c_{j,s'}\\
    & + i \lambda_{soc} \sum_{\langle \langle ij\rangle \rangle} c^\dagger_{i,s} \nu_{ij}  \sigma^z_{s,s'} c_{j,s'},
    \end{aligned}
    \end{equation}
    where $c^\dagger_{i,s}$ represents a creation of Wannier state at a cluster $i$ with spin $s$, and ${\vec \sigma}$ is a Pauli matrix for spin-1/2. $n_{i,s}=c^\dagger_{i,s} c_{i,s}$.
    ${\hat d}_{ij}$ is the unit vector from site $j$ to site $i$ introduced due to the broken inversion symmetry and $\nu_{ij}$~=~$\pm$1 depends on whether the electron hoping from $i$ to $j$ makes a right (+1) or a left (-1) turn.
    $\langle i,j \rangle $ and $\langle \langle i,j \rangle \rangle$ refer to first and second nearest neighbor (n.n.) respectively, and hopping parameters $t_i$ up to 5th n.n. are used to fit the band dispersion. 
    $\lambda_{soc}$ and $\lambda_R$ denote strengths of spin-dependent and Rashba spin-flip SOC, respectively.
    Tight binding parameters are listed in Table~\ref{table1}, which fit the DFT bands well (Fig.~3(c)).
    The calculated Wannier charge center~\cite{Soluyanov2011} indicates clearly an odd number of crossing via the pumping parameter, a hallmark of nontrivial topology of Z$_2$=1 (Fig.~3(g)).
    The nontrivial order can also be manifested by the presence of helical edge states (see SFig.~7 in the Supplemental Material~\cite{supplement}).

%%%%%%%%%%%%%%%%%%%%%%%%%%%%%%%%%%%%%%%%%%%%%%%%%%%%%%%%%%%%%%%%%%%%%%
\begin{table} [ht]
\caption{\label{table1}Parameters of the tight binding Hamiltonian of K-adsorbed 1\textit{T}-TaS$_2$ at the optimal coverage.}
\centering
\begin{tabular}{cccc}

\hline
\hline
Parameter  & Value (meV) & Parameter & Value (meV) \\
\hline
t$_{1}$ &  $-$0.67 & t$_2$ & $-$1.06  \\
t$_{3}$ &  $-$1.32 & t$_4$ & 0.26  \\
t$_{5}$ &  0.62 & $\lambda_R$ & 0.07  \\
$\lambda_{soc}$ &  0.06 & & \\
\hline
\hline
\end{tabular}
\end{table}
%%%%%%%%%%%%%%%%%%%%%%%%%%%%%%%%%%%%%%%%%%%%%%%%%%%%%%%%%%%%%%%%%%%%%%

\textit{Electron correlation and honeycomb-lattice Mott insulator}---While the DFT results capture the narrow Ta~5$d$ bands, its tiny gap is inconsistent with the 200~meV gap observed (see Fig.~2(g) and SFig.~8 in the Supplemental Material~\cite{supplement}).
    This suggests a crucial role of electron-electron interactions.
    The electron correlation is considered by performing DMFT calculation using
    the above Wannier-function tight-binding parameters.
    As shown in Fig.~3(h), this calculation yields the fully gaped band structure even with a fairly small Coulomb interaction of 0.25~eV due to the very small hopping integrals or the narrow bands.
    This unambiguously indicates the strongly correlated, Mott insulating, nature of the present system. 
    The occupied and unoccupied bands around -0.1 and 0.1 eV form the LHB and UHB, respectively.
    The band gap and the narrow LHB band agree well with the above STS results and our angle-resolved photoemission spectroscopy measurements (see SFig.~8 in the Supplemental Material~\cite{supplement}).
   The topological character of DFT metallic bands (Fig.~2(b)) is preserved here.
   For example, the LHB and UHB are made of two bands with opposite Chern numbers.
   When the system is half filled as in the current system, it is a Mott insulator
with zero Chern number because two Chern numbers cancel out.
On the other hand, at 1/4 and 3/4 fillings, it becomes a quantum spin Hall insulator.
This is apparently a Kane-Mele-Hubbard system.

Another possible origin of the insulating phase in our system is a magnetic ordering. Any magnetic orderings in a honeycomb lattice drive the insulating phase due to the time-reversal-symmetry breaking. We also considered various magnetic orderings through DFT and spin model calculations but the transition temperature is extremely small ($\sim$ $O(mK)$)~\cite{supplement}. Therefore, the experimentally observed charge gap, persisting well above such a low transition temperature (see SFig.~8 in the Supplemental Material~\cite{supplement}), is a strong evidence of the insulating phase without any symmetry breaking, that is the Mott insulator.

    We note that the direct experimental confirmation of the topological property itself is missing in the parent work.
    Given that two propagating edge states have different Chern numbers, there is no edge states in half-filled system.
    The verification of the topological character, thus, requires doping the system to either 1/4 or 3/4 filling, or removing one of edge states by applying magnetic field to induce a finite Hall conductivity originated from the left-over edge state. The quantized Hall conductivity and STS in a magnetic field in a single layer of K-absorbed 1\textit{T}-TaS$_2$ would be excellent topics of further works together with the tuning of the electron filling to find various exotic phases predicted theoretically. A very recent theoretical work suggested the realization of a buckled honeycomb lattice in 1\textit{T}-TaSe$_2$ through a particular stacking of two layers \cite{Pizarro}. This system, however, is not expected to be within the Mott insulator regime due to a substantially, two orders of magnitude, smaller $U$/$t$ value \cite{Pizarro}.

\textit{Conclusion}---In this work, we create a honeycomb lattice of half-filled $d$ electrons on a 1{\textit {T}}-TaS$_2$ surface, which is decorated with ordered K adsorbates. 
  The STS and ARPES data and theoretical considerations indicate unambiguously that the finite-temperature insulating phase of the present system is a honeycomb-lattice Mott insulator described by a supercell Kane-Mele-Hubbard model with a finite spin Chern number for lower and upper Hubbard bands. 
The lattice manipulation by adsorbate superstructures can widely be exploited in various other 2D materials, in addition to moire superstructures, in order to search and create an exotic electronic system.\\
\\

\textit{Acknowledgments.}---This work was supported by the Institute for Basic Science (Grant No. IBS-R014-D1).
K.-H.~Jin and F.~Liu acknowledge financial support from DOE-BES (No. DE-FG02-04ER46148).
K.-H.~Jin is supported by the Institute for Basic Science (Grant No. IBS-R014-Y1).
A.~Go thanks financial support from the Institute for Basic Science (Grant No. IBS-R024-D1).
H.-Y.~Kee is supported by Natural Science and Engineering Research Council of Canada.

\bibliographystyle{apsrev4-1}
\bibliography{K_1T-TaS2_PRL.bib}

\end{document}